# Notch Fracture predictions using the Phase Field method for Ti-6Al-4V produced by Selective Laser Melting after different post-processing conditions

A. Díaz[1]*, J.M. Alegre[1], I.I. Cuesta[1], E. Martínez-Pañeda[2], Z. Zhang[3]

[1]Universidad de Burgos. Escuela Politécnica Superior, Avda. Cantabria s/n, 09006, Burgos, Spain

[2]Department of Civil and Environmental Engineering, Imperial College London, London SW7 2AZ, UK

[3]NTNU Nanomechanical Lab, Department of Structural Engineering, Norwegian University of Science and Technology (NTNU), Trondheim 7491, Norway

*Contact author: adportugal@ubu.es

**Abstract**

Ti-6Al-4V is a titanium alloy with excellent properties for lightweight applications and its production through Additive Manufacturing processes is attractive for different industrial sectors. In this work, the influence of mechanical properties on the notch fracture resistance of Ti-6Al-4V produced by Selective Laser Melting is numerically investigated. Literature data is used to inform material behaviour. The as-built brittle behaviour is compared to the enhanced ductile response after heat treatment (HT) and hot isostatic pressing (HIP) post-processes. A Phase Field framework is adopted to capture damage nucleation and propagation from two different notch geometries and a discussion on the influence of fracture energy and the characteristic length is carried out. In addition, the influence of oxygen uptake is analysed by reproducing non-inert atmospheres during HT and HIP, showing that oxygen shifts fracture to brittle failures due to the formation of an alpha case layer, especially for the V-notch geometry. Results show that a pure elastic behaviour can be assumed for the as-built SLM condition, whereas elastic-plastic phenomena must be modelled for specimens subjected to heat treatment or hot isostatic pressing. The present brittle Phase Field framework coupled with an elastic-plastic constitutive analysis is demonstrated to be a robust prediction tool for notch fracture after different post-processing routes.

## 1. Introduction

The adoption of Additive Manufacturing (AM) techniques as a competitive alternative to conventional production of industrial components still poses some important challenges; despite AM being cost-effective for complex or customised geometries, the need for post-processing operations increases production times and thus limits the efficiency of the process. In the present work, a numerical framework based on the Phase Field model is presented to predict fracture of additively manufactured Ti-6Al-4V after different post-processing conditions and under different stress concentration states.

Layer-upon-layer fabrication alters mechanical behaviour for both polymer [1] and metallic [2] additively manufactured parts. For instance, fracture of polymer samples produced by Fused Deposition Modelling is governed by fibre debonding and is thus sensitive to printing patterns [3]. On the other hand, Powder Bed Fusion techniques for metallic AM, e.g. Selective Laser Melting (SLM), are characterised by a strong thermal gradient during printing, which produces residual stresses and usually hard and brittle microstructures [4], requiring thus a post-processing stage [2] to relieve the residual stress state and to maximise toughness.

Titanium alloys have been extensively applied in the aerospace and biomedical fields where a high strength-to-weight ratio is required. Additionally, AM methods are suitable due to the need for special geometries and short-run production for these applications [5]. Ti-6Al-4V, the most common Titanium alloy, is a dual $\alpha + \beta$ alloy that combines a high yield stress with a good toughness and corrosion resistance [6]. This alloy has been extensively studied considering AM methods such as Selective Laser Melting (SLM) [7,8], Electron Beam Melting (EBM) [9], Wire Arc AM (WAAM) [10] or other direct deposition methods [11]. EBM can produce Ti-6Al-4V parts at high build rates but produces poor surface finish [12,13]; on the other hand, SLM is characterised by thermal gradients and rapid cooling resulting in a martensitic $\alpha'$ transformation [14]. Therefore, SLMed Ti-6Al-4V specimens show a high strength but low ductility and require a post heat treatment [15] that softens the microstructure and thus improves toughness. To reduce the size of defects after SLM and increase fatigue life, a Hot Isostatic Pressing (HIP) procedure is commonly carried out [16,17].

During the thermal post-processing of titanium alloys, oxygen uptake is a great concern since ductility is critically reduced with oxygen content [18] and hence heat treatments or HIP should be performed in inert atmospheres. Despite the use of Argon environments for heat treatments, oxygen impurities produce alpha case layers, even in high-purity Ar and for only 0.2 ppm of $O_2$ [19]. Vacuum atmospheres can also be used to prevent the formation of alpha case [19]. The kinetics of alpha case formation are rate-controlled by oxygen diffusion through the $\beta$ bcc phase. Therefore, the most common strategies for reducing the formation of the alpha layer are: (i) reduction of oxygen content in the environment, and (ii) acceleration of cooling to minimise times at high temperatures. The identification of the alpha case thickness can be established from a micro-hardness or concentration profile, e.g. through EPMA measurements [20], but limiting values of HV or concentration should be specified. Richards [21] defines an "effective alpha case" as the layer in which brittle fracture is found, as shown in Figure 1; this depth depends on the testing temperature and not only on the concentration profile.

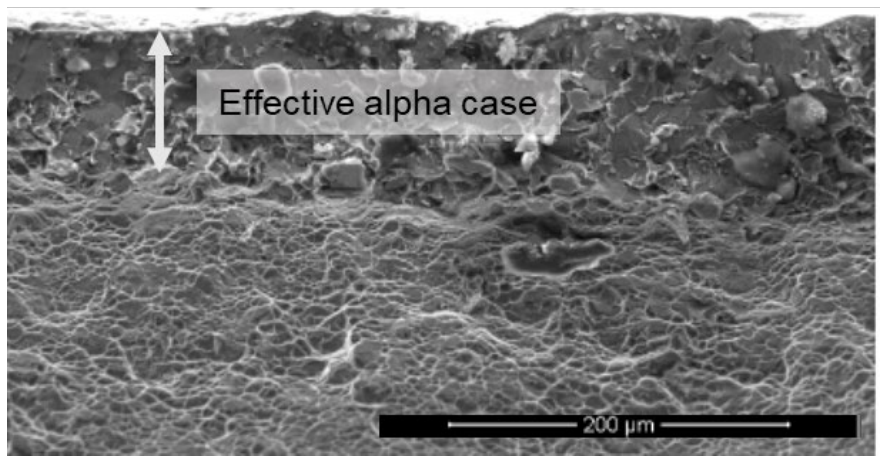

*Figure 1. Effective alpha case for tensile testing of fine-grained Ti-6Al-4V [21].*

To sum up, thermomechanical treatment of Ti-6Al-4V can result in a profound modification in microstructure and thus on mechanical properties, especially in fracture toughness and fatigue resistance [16]. This wide behaviour range is challenging for fracture modelling as requires encompassing both brittle and ductile modes of failure.

Notch fracture analysis usually relies on elastic assumptions, i.e. it is based on Linear Elastic Fracture Mechanics (LEFM), which can be valid for the brittle response of SLMed Ti-6Al-4V but cannot be assumed after the toughening effect of post-processing.

For instance, the Theory of Critical Distances (TCD) has proven to be a simple and robust methodology to predict notch fracture [22,23] by finding an inherent strength that can be related to the ultimate tensile stress in notched samples. However, for softened microstructures of Ti-6Al-4V, which can attain very high $K_C$ values [16], elastic assumptions for the TCD calculations are not valid for fracture predictions.

On the other hand, damage modelling has exploited the growing computational capabilities of Finite Element analysis, and many approaches are now available to simulate crack nucleation and propagation. Continuum damage modelling avoids displacement discontinuities and thus overcomes issues intrinsic to discrete approaches, but constitutive softening leads to strain localisation and thus regularisation or non-local strategies must be considered to circumvent mesh dependency [24]. Within this context, Phase Field (PF) models have been become popular due to their versatility in predicting damage nucleation, propagation and branching [25,26]. PF models for brittle fracture extend Griffith's theory to a damage framework in which crack nucleation and propagation are simulated considering a diffusive crack topology [27], as detailed in Section 3.2. The crack discontinuity is thus smeared over a band proportional to a characteristic length scale. Ductile modifications for the Phase Field framework have been proposed by including a plastic energy density term in the energy functional and by modifying the degradation function [28]. Other authors have considered elastic-plastic undamaged behaviour in combination with a brittle Phase Field model governed purely by elastic strains [29]. Similarly, Kristensen et al. [30] implemented strain gradient plasticity in a numerical framework for hydrogen embrittlement modelling, and these authors also neglected the plastic contribution in the Phase Field local balance. Martínez-Pañeda et al. [31–34] have adapted Phase Field models to predict hydrogen assisted cracking through the reduction of the critical energy as a function of local hydrogen concentration. That mechanical framework informed by diffusion analysis inspires the present oxygen-enhanced embrittlement modelling, as shown in Section 3. However, the present modelling is simpler since oxygen is assumed to be immobile during room temperature tensile testing.

In order to assess the influence of fracture toughness increase on notch mechanics within a Phase Field framework, three conditions are extracted from literature for as-built SLM, heat treated and HIPed specimens, as described in Section 2. The informed Phase Field framework is presented in Section 3, where the oxygen diffusion and modification of mechanical properties are also detailed. Results for two notch shapes are discussed in terms of the characteristic length choice and considering elastic or elastic-plastic behaviour. In addition, the influence of thermo-mechanical treatments under a hypothetical non-inert atmosphere is reproduced to account for oxygen embrittlement.

## 2. Simulated conditions and geometry

The study of notch fracture behaviour of as-built SLM and the effect of post-processing is performed considering the tensile and toughness properties experimentally found by Zhang et al. [16] in Ti-6Al-4V samples. In that work, fracture toughness was determined in 5-mm Compact Tension specimens and $K_C$ in Table 1 represents a plane-stress toughness. Three conditions were compared:

- As-built condition (SLM). After Selective Laser Melting, the mean value of fracture toughness is only 54.6 MPa·m$^{0.5}$; this brittle behaviour is caused by the martensitic microstructure and partly attributed to residual stress.
- Heat treated condition (HT). SLMed samples were subjected to a thermal post-processing step at 780ºC for 4 hours in vacuum, resulting in a great increase in fracture toughness.
- Hot Isostatic Pressing condition (HIP). SLMed samples were HIPed at 940ºC and 125ºC in an Argon atmosphere. Despite the coarser $\alpha$ width in the HIP condition relative to the heat treated samples, an unexpected slightly lower toughness was found, which is attributed to the experimental scatter.

Tensile properties were slightly affected by post-processing, with the HIP condition resulting in the lowest yield stress. The hardening coefficient $n$ has been fitted here assuming the common relationship $(n/0.002)^n = \sigma_u/\sigma_{ys}$ [35], where $\sigma_u$ is the ultimate tensile strength and $\sigma_{ys}$ the yield stress.

*Table 1. Mechanical properties for the as-built SLM and for HT and HIP post-processing conditions. Average results extracted from [16].*

|  | SLM | HT | HIP |
|---|---|---|---|
| $E^0$ (GPa) | 103.6 | 113.0 | 107.1 |
| $\sigma_{ys}^0$ (MPa) | 810.6 | 821.5 | 762.9 |
| $n$ | 0.056 | 0.035 | 0.041 |
| $K_C^0$ (MPa·m$^{0.5}$) | 54.6 | 145.4 | 134.1 |

From the experimental findings of Zhang et al. [16] regarding the post-processing influence on fracture toughness, notch behaviour is numerically analysed considering two different stress concentrators: a V and a C-shaped notches. These geometries are extracted from Peron et al. [23], who exploited the difference in stress profiles to apply the Theory of Critical Distances (TCD) and predict notched fracture. The thickness of the simulated specimens is 3 mm and notch dimensions are shown in Figure 2.

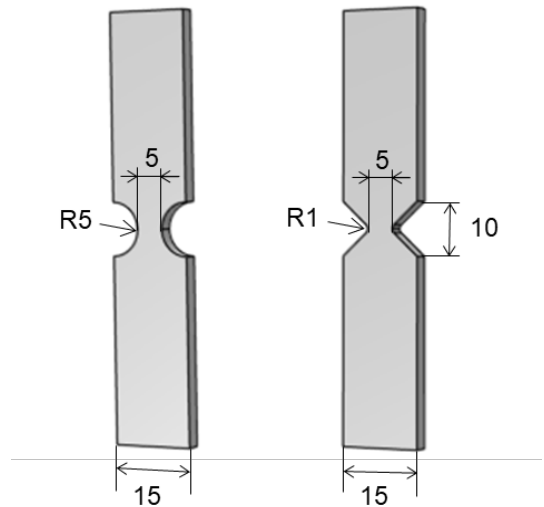

*Figure 2. Geometry of C-notched and V-notched tensile specimens (dimensions in mm).*

In the original work of Zhang et al. [16], the effect of machining was assessed but was found to have a negligible effect on tensile properties and fracture toughness. Therefore, it is concluded that oxygen uptake did not occur in the vacuum (HT) or Argon (HIP) atmospheres. With the aim of analysing a hypothetical oxygen impurity fraction at these temperatures and pressure conditions, two scenarios are here simulated as shown in Table 2: for HT post-processing, conditions of laboratory air are simulated, i.e. a 21% of molar fraction, whereas for HIP, a low oxygen fraction is reproduced to account for possible Argon contamination.

Table 2. Conditions for the simulated heat treatment (HT) and hot isostatic pressing (HIP).

|  | HT | HIP |
|---|---|---|
| Temperature (°C) | 780 | 940 |
| Pressure (MPa) | 0.1 | 125 |
| Holding time (h) | 4 | 1.5 |
| Inert atmosphere [16] | Argon | Vacuum |
| Simulated O$_2$ fraction for non-inert conditions | 0.21 | 0.0001 |

## 3. Numerical modelling

As mentioned before, the numerical framework includes an oxygen diffusion step that informs the subsequent Phase Field modelling. The local reduction in fracture toughness is inspired by previous numerical studies on hydrogen embrittlement [31]. However, oxygen can be assumed to be immobile during testing at room temperature and thus diffusion equations do not need to be solved during the simulation of damage evolution. Both physical problems are simulated using the Finite Element software Comsol Multiphysics 6.0.

### 3.1. Oxygen diffusion

Thermal oxidation in titanium alloys and the subsequent oxygen diffusion from the oxide/metal interface is a complex phenomenon that requires the combined modelling of oxidation kinetics and interstitial diffusion. Wagner's model proposes a linear distribution of oxygen concentration in the oxide phase since the solubility range in TiO$_2$ is small [36], and the analytical solution of Fick's laws can be considered for the interstitial oxygen within the bulk metal that forms an "oxygen-enriched layer" (OEL).

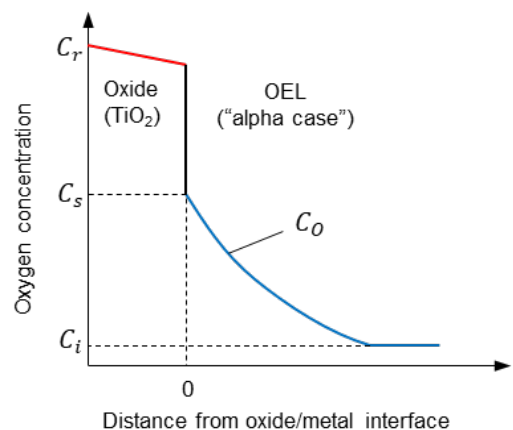

Figure 3. Scheme of oxide layer and oxygen concentration profiles (adapted from [36]).

Oxide growth in Ti-6Al-4V is even more complex due to the intercalated formation of aluminium oxides [37]. Nevertheless, the size of the oxide layer is commonly much smaller than the oxygen-enriched layer and thus the concentration at the oxide/metal interface, i.e. $C_s$ in Figure 3, is assumed here as the boundary condition for the diffusion problem.

Diffusion of interstitial oxygen is simulated in the Comsol Multiphysics module "Transport of Diluted Species", where the governing equation for the diffusing species represents a mass balance, i.e. Fick's second law:

$$\frac{\partial C_O}{\partial t} - \nabla \cdot (D_O \nabla C_O) = 0 \qquad (1)$$

where $C_O$ is the concentration of oxygen in solid solution and $D_O$ the diffusivity value at the corresponding temperature. An Arrhenius law is assumed for the increase in diffusivity with temperature:

$$D_O = D_O^0 \exp\left(-\frac{E_a}{RT}\right) \qquad (2)$$

where $D_O^0$ is the pre-exponential factor, $E_a$ the activation energy for lattice diffusion, $T$ the temperature and $R$ the constant of gases. Even though the concentration at the oxide/metal interface is governed by oxidation kinetics, oxygen uptake at the surface depends on oxygen partial pressure [38]. To account for that effect a simplified approach is considered here and a Sievert-type expression is assumed for modelling boundary conditions of the oxygen diffusion problem even though $O_2$ dissociation and adsorption are not occurring in that interface. Thus, the concentration at the boundary, $C_s$ is expressed as follows:

$$C_s = K_O \sqrt{p_{O_2}} \qquad (3)$$

where $K_O$ represents oxygen solubility and $p_{O_2}$ the oxygen partial pressure. Solubility is also assumed to follow an Arrhenius law, with a pre-exponential factor $K_O^0$ and an activation energy for solubility $E_s$:

$$K_O = K_O^0 \exp\left(-\frac{E_s}{RT}\right) \qquad (4)$$

From the experimental results of Casadebaigt et al. [20] at 500ºC and 600ºC, diffusivity and solubility parameters have been fitted (Table 3). The use of the extrapolated values to higher temperatures is an oversimplification but it is here assumed to illustrate the modelling capabilities. The relationship between the partial pressure of oxygen pressure and the total atmospheric pressure is here assumed to follow Dalton's law, $p_{O_2} = x_{O_2} p$, where $x_{O_2}$ is the mole fraction of oxygen.

Table 3. Properties for oxygen uptake and diffusion, fitted from [20].

| | |
|---|---|
| $D_O^0$ (m²/s) | 4.0×10⁻⁴ |
| $E_a$ (kJ/mol) | 219.0 |
| $K_O^0$ (wt.%/MPa^0.5) | 123.86 |
| $E_s$ (kJ/mol) | 53.1 |

### 3.2. Phase Field modelling

A Solid Mechanics analysis is carried out in Comsol Multiphysics considering a linear elastic material and also a classical J2 plasticity with an isotropic power-law hardening when the elastic-plastic behaviour is assumed:

$$\sigma_{ys} = \sigma_{ys}^0 \left(1 + \frac{E}{\sigma_{ys}^0} \varepsilon^p\right)^n \quad (5)$$

where $\sigma_{ys}^0$ is the initial yield stress, $n$ the hardening exponent, as defined in Table 1, $E$ the Young's modulus and $\varepsilon^p$ the equivalent plastic strain. This hardening behaviour is implemented in Comsol Multiphysics through the available Swift's model for plasticity. There are three strain energy densities in the problem: the elastic strain energy densitiy $\psi_e$, the dissipative plastic strain energy density $\psi_p$ and crack surface energy density $\psi_s$ [30]:

$$\psi = \psi_e + \psi_p + \psi_s \quad (6)$$

For the elastic material scenario, i.e. without plasticity, Phase Field modelling is driven only by elastic strains in the present work. However, when elastic-plastic behaviour is simulated the dissipated plastic energy is included in the Phase Field equilibrium, as detailed below. However, damage and plasticity are decoupled since the plastic term $\psi_p$ is independent of the phase field $\phi$ whereas the degradation function $g(\phi)$ does not consider a plastic influence [39]. Hence, the elastic energy density is decomposed into a traction $\psi_e^+$ and compression $\psi_e^-$ parts, assuming that only the traction term contributes to the crack propagation:

$$\psi_e = g(\phi)\psi_e^+ + \psi_e^- \quad (7)$$

The dissipated energy caused by plastic deformation is calculated in Comsol Multiphysics by integrating the rate of plastic energy density:

$$\dot{\psi}_p = \boldsymbol{\sigma}:\dot{\boldsymbol{\varepsilon}}_p \quad (8)$$

where $\boldsymbol{\sigma}$ and $\dot{\boldsymbol{\varepsilon}}_p$ represent the stress and plastic strain rate tensors, respectively. The dissipated $\Psi_p$ is accessed in Comsol Multiphysics as the variable *solid.Wp*.

The smeared crack that characterises Phase Field modelling emerges from the following approximation of the fracture surface energy that adopts a diffusive crack topology over a $l_c$ width:

$$\psi_s = G_c \left(\frac{\phi^2}{2l_c} + \frac{l_c}{2}\nabla^2\phi\right) \quad (9)$$

where $G_c$ represents the critical strain energy release rate or Griffith's fracture energy. To simulate Phase Field fracture, an in-built capability is available in the Comsol

Multiphysics 6.0 version within the module of Nonlinear Structural Mechanics. However, that analysis is not compatible with plasticity, and thus an alternative implementation is performed following the work of Zhou et al. [40]. The most common degradation function is considered:

$$g(\phi) = (1-k)(1-\phi)^2 + k \qquad (10)$$

where $k$ is a numerical parameter taking a very small positive value ($0 < k \ll 1$) to prevent numerical singularities; here: $k = 10^{-6}$. Imposing a phase field local balance, the governing equation driving damage is implemented in Comsol Multiphysics through a Helmholtz equation since it can be expressed as [40]:

$$\left[\frac{2l_c(1-k)H}{G_c} + 1\right]\phi - l_c^2 \nabla^2 \phi = \frac{2l_c(1-k)H}{G_c} \qquad (11)$$

where $H$ is a history variable that represents the maximum energy density that has been reached at each point and includes also the plastic dissipative term, i.e. $H = \max(\psi_e^+ + \psi_p)$; this replacement the current energy density by the strain-history variable $H$ ensures the irreversibility of phase field damage. The stress calculation considering the degraded material is carried out through a user-defined elasticity matrix; more details can be found in [40]. In the present work, the degradation of the elasticity matrix needs to be redefined in terms of elastic principal strains, in contrast to the original implementation of Zhou et al. [40], and the diagonalisation of the elastic strain matrix is required in Comsol Multiphysics. Geometrical nonlinear effects are not considered here and small strains are assumed in the plastic model.

The choice of the parameters $G_c$ and $l_c$ determines the fracture response: low $G_c$ or high $l_c$ values produce a shift to brittle failures. The influence of the characteristic length $l_c$ is assessed while the critical Griffith's energy $G_c$ is here fixed with a value derived from the plane-stress toughness from Zhang et al. [16]:

$$G_c = K_c^2/E \qquad (12)$$

where $E$ is Young's modulus. It must be noted that both $G_c$ and $l_c$ influence the critical stress $\sigma_c$ in the cohesive crack behaviour that takes place locally during damage evolution. Borden et al. [41] derived the following relationship:

$$\sigma_c = \frac{9}{16}\sqrt{\frac{EG_c}{3l_c}} \qquad (13)$$

Even though that relationship was derived for a one-dimensional configuration, it illustrates the opposite effect of $G_c$ and $l_c$ on fracture. In addition, the length scale determines mesh density : different element sizes have been simulated in order to discard possible strain localisation mesh effects in elastic-plastic models, but an element size of $l_e = l_c/5$ in the whole notch ligament is fixed for all the present simulations showing the exact same results than a finer mesh ($l_e = l_c/7.5$) but with a lower computational cost. As demonstrated by other authors [42], this ensures that phase field evolution is correctly captured within the fracture process zone and results are mesh-insensitive. Once it is verified that fracture does not change between $l_e = l_c/5$ and $l_e = l_c/7.5$ the finer mesh is applied to the notch surface to better capture the oxygen concentration profile. Remeshing capabilities are not exploited here but some authors have shown a substantial reduction in the computational cost with remeshing [43] or even with neural network techniques that avoid discretization [44].

When elastic-plastic modelling is considered, plastic strain accumulation can lead to damage due to the effect of dissipative terms in the crack driving force. Therefore, without any remeshing algorithm, the mesh needs to be refined not only near the notch plane but in a relatively wide strip. For the C and V-notch, a layer of 3 and 1.5 mm of width and centred on the notch plane is finely meshed, i.e. with $l_e = l_c/5$, when plastic effects are incorporated.

To resolve the Solid Mechanics problem, a segregated approach is considered in which the degrees of freedom for displacements are solved in an iterative scheme separated from the Phase Field variables, i.e. $\phi$ and $H$. The strain-history variable, in contrast to the work of Zhou et al. [40], is implemented as a state variable. This segregated step approach subdivides the coupled problem and solves sequentially all the couplings between displacements and $\phi$ in each iteration, demanding less memory than a fully coupled scheme [45]. In the present work, a stationary step is chosen for the Solid Mechanics problem, and loading is modelled using a parametric sweep with a displacement increment of 0.002 mm and a solver tolerance of $10^{-4}$.

### 3.3. Oxygen-modified mechanical properties

For the hypothetical oxygen conditions shown in Table 2, the influence of the alpha case formation on notch fracture resistance is assessed. For that purpose, a degradation law is assumed but, in contrast to the aforementioned works on hydrogen embrittlement, a linear reduction is assumed here for $K_c$:

$$K_c = K_c^0(1 - \chi C_O[\text{wt.\% O}]) \qquad (14)$$

where $K_c^0$ is the fracture toughness in the oxygen-free condition, taking the values of Table 1, and the dimensional coefficient $\chi$, in [wt.%]$^{-1}$, quantifies embrittlement. The assumption of linear reduction in $K_c$ is based on the fact that Young's modulus is strongly influenced by oxygen concentration so a linear degradation on $G_c$ could not reproduce embrittlement at high concentrations. Following the experimental results from Lee [46]:

$$E(C_O) = E^0 \left(1 + \frac{13.5[\text{GPa}]}{E^0} C_O[\text{wt.\% O}]\right) \qquad (15)$$

From equations (12), (14) and (15), the evolution of $G_c$ with increasing oxygen concentration is plotted in Figure 4, considering an embrittlement coefficient $\chi = 0.01$. When the elastic-plastic behaviour is assessed, the oxygen-enhanced hardening is also reproduced and a value of $\xi = 0.1$ is assumed as a reasonable trend [47]:

$$\sigma_{ys} = \sigma_{ys}^0(1 + \xi C_O[\text{wt.\% O}]) \qquad (16)$$

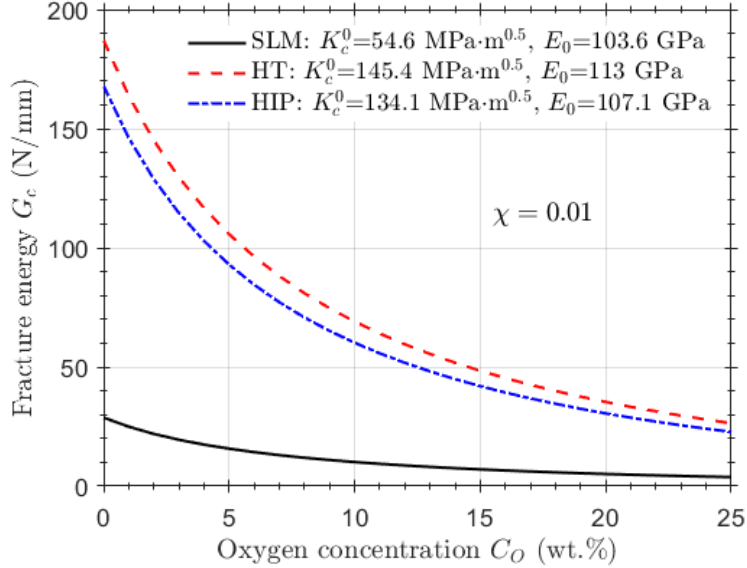

Figure 4. Reduction in fracture energy as a function of oxygen concentration.

## 4. Results and discussion

### 4.1. Length scale influence

The whole ligament must be finely meshed to capture the final failure, especially for the elastic-plastic conditions in the C-notch where damage is accumulated in the centre of the specimen. After a mesh sensitivity study, it has been found that a larger band must be meshed finely in the C-notch due to the lower stress concentration. In addition, the smaller $l_c$ results in a higher number of elements since the element size is fixed as $l_e = l_c/5$ within the notch ligament. Therefore, larger computational resources are required to solve the Phase Field – Displacement numerical problem for the situation with $l_c = 0.05$ mm. The number of elements for each condition is shown in Table 4.

Table 4. Number of elements for the evaluated notch geometries and characteristic lengths.

|         | $l_c = 0.05$ mm | $l_c = 0.10$ mm | $l_c = 0.30$ mm |
|---------|-----------------|-----------------|-----------------|
| C-notch | 314,226         | 91,460          | 10,592          |
| V-notch | 68,664          | 18,420          | 3,096           |

Figures 5 and 6 demonstrate that a smaller characteristic length results in fracture at higher forces and lower displacements. All fracture curves show the reaction force in the upper surface against the applied displacement. This $l_c$ effect in Phase Field fracture modelling has been previously demonstrated by other authors and is anticipated by the relationship of this value with a local critical stress, as expressed in equation (13). For a bigger length scale, e.g. $l_c = 0.3$ mm, the crack propagates over a broader band due to the $l_c$ control over the crack diffusive topology and this results in a more brittle behaviour in comparison to $l_c = 0.1$ mm. In order to assess whether the critical stress $\sigma_c$ from equation (13) is the sole parameter influencing fracture, the fracture energy $G_c$, corresponding to the as-built SLM condition, is divided by 3 while fixing $l_c = 0.1$ mm. However, this combination of parameters does not match the force-displacement curve of $G_c$ and $l_c = 0.3$ mm, indicating that fracture behaviour does not only depend on the ratio $G_c/l_c$.

The simulated tensile behaviour of the C-notch geometry (Figure 5) shows failures at higher maximum forces and displacements in comparison to the V-notch (Figure 6) with the same characteristic length, as expected due to the stronger stress concentration in the V-notch geometry.

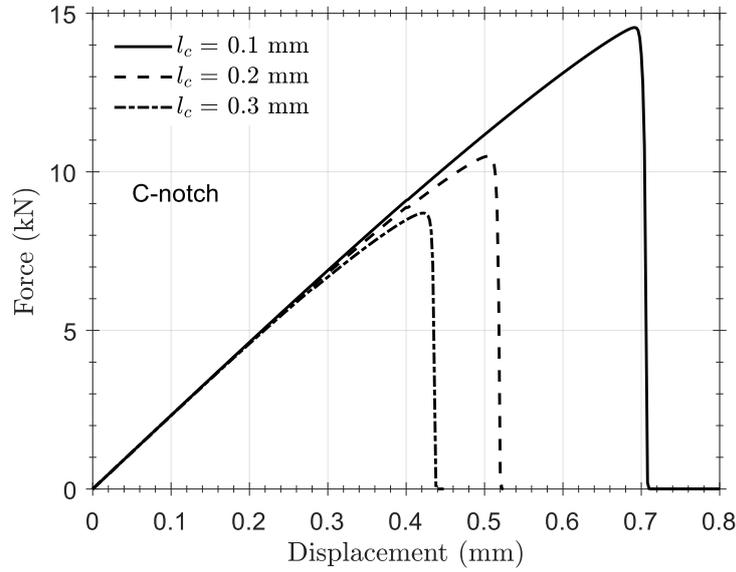

*Figure 5. $l_c$ influence on C-notch fracture with elastic behaviour.*

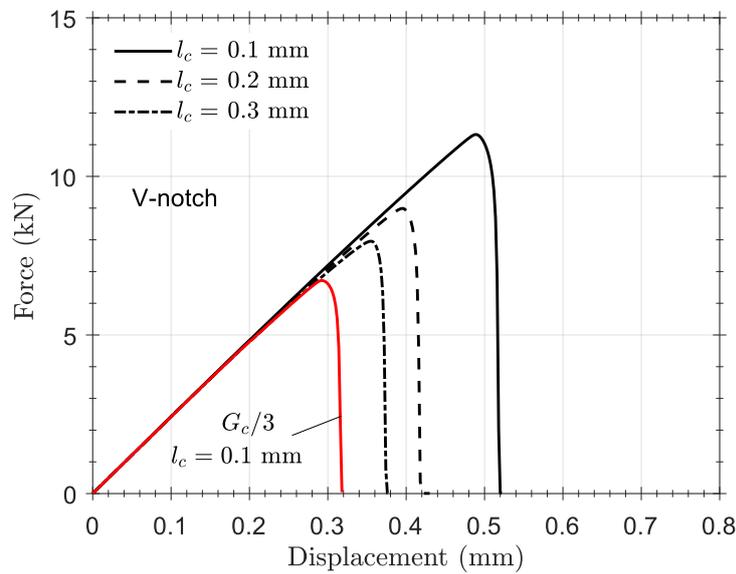

*Figure 6. $l_c$ influence on V-notch fracture with elastic behaviour.*

Unstable crack propagation begins when the phase field $\phi$ reaches a value close to 1.0, as shown in Figure 7; however, before that event, damage is spread over a bigger region around the notch tip. When only elastic behaviour is reproduced, crack propagation occurs abruptly, and small displacement increments must be employed to capture the steep drop in force observed. In this work, displacement increments of 0.002 were used.

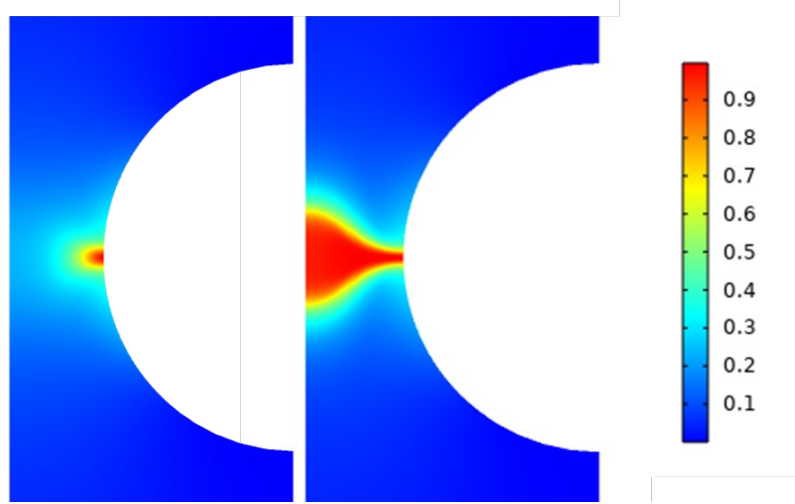

*Figure 7. Phase Field ($\phi$) at crack initiation (left) and at final failure (right) for SLM, $l_c =$ 0.1 mm and elastic behaviour.*

Force values reached for the linear elastic behaviour are not realistic for a Ti-6Al-4V sample until characteristic lengths higher than 0.3 mm are chosen. However, when elastic-plastic material behaviour is modelled, the stress level is not governed by $l_c$; the phase field length scale magnitude mainly influences the maximum displacement before fracture, as shown in Figures 8 and 9 for the C-notch and V-notch, respectively. Regarding the mode of failure, it is observed that plasticity shifts the damage nucleation site from the notch surface to the specimen centre for the C-notch geometry, as shown in Figure 10. This phenomenon is common for low triaxiality states, where plastic strain accumulates at the specimen centre and triggers ductile failure. It can be concluded that for a big notch radius the failure mode is not governed by fracture near the notch tip, but by a damage accumulation in the central section of the specimen, as experimentally show in [48] for hydrogen embrittlement studies on notched specimens. Since the plastic dissipative part is included in the $H$ variable accounting for the driving strain energy, damage nucleates in highly plasticised regions, as demonstrated by previous authors [28], and damage shear bands are observed at the final failure in Figure 10 for the C-notch.

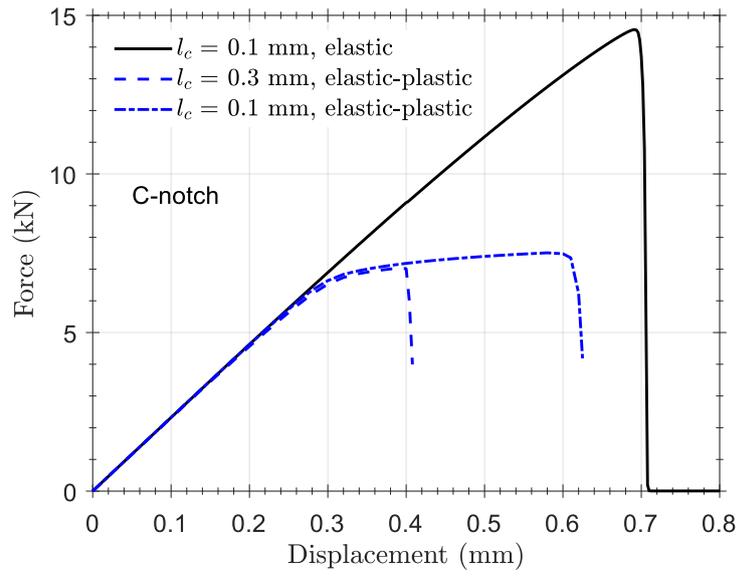

Figure 8. Plasticity influence on C-notch fracture.

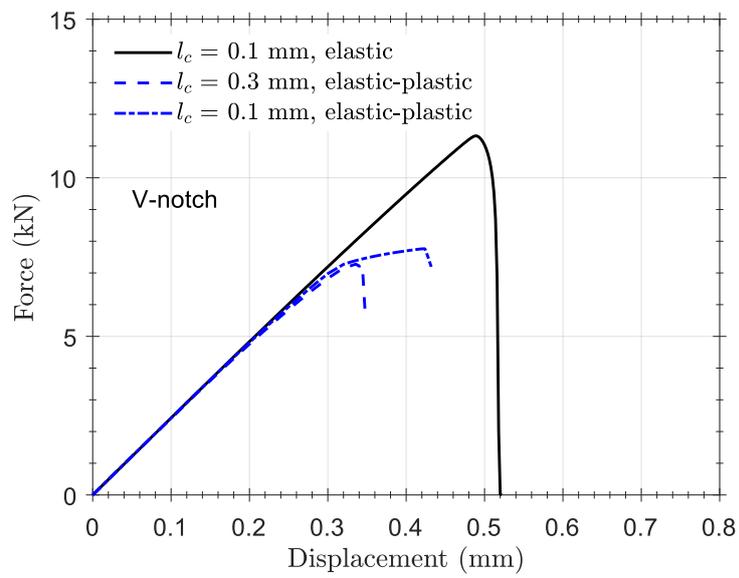

Figure 9. Plasticity influence on V-notch fracture.

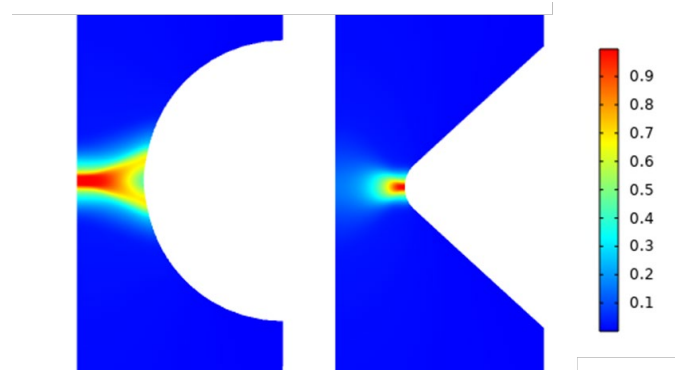

*Figure 10. Phase Field (ϕ) at final failure for SLM, $l_c = 0.1$ and elastic-plastic behaviour in both notch geometries.*

### 4.2. Post-processing influence

As expected, the increase in fracture energy $G_c$ from the 28.8 N/mm in the as-built SLM condition to 167.9 and 187.1 N/mm after the HIP and HT post-processing, respectively, leads to a very ductile behaviour. If a purely elastic material is considered, notch fracture for both C and V-notches takes place at unrealistic force values for $l_c = 0.1$ mm, as shown in Figures 11 and 12, equivalent to engineering stresses higher than 5000 and 6000 MPa. The brittle martensitic $\alpha'$ microstructure in the as-built SLM samples is transformed into acicular $\alpha$ and the residual stress level is expected to be reduced after heat treatment (HT) [16]; on the other hand, the higher temperature during HIP causes an increase in grain size and in the $\beta$-Ti phase. Therefore, HT and HIP conditions cannot be modelled assuming a linear elastic and brittle behaviour, and plasticity should be included to capture the significant elongation level before fracture. This shift to ductile behaviour is observed in Figures 12 and 13 for the C-notch and V-notch geometries when J2 classical plasticity with an isotropic power-law hardening has been considered, even though the Phase Field approach is elastic, i.e. the governing balance does not consider coupled behaviour of plastic terms and crack surface energy.

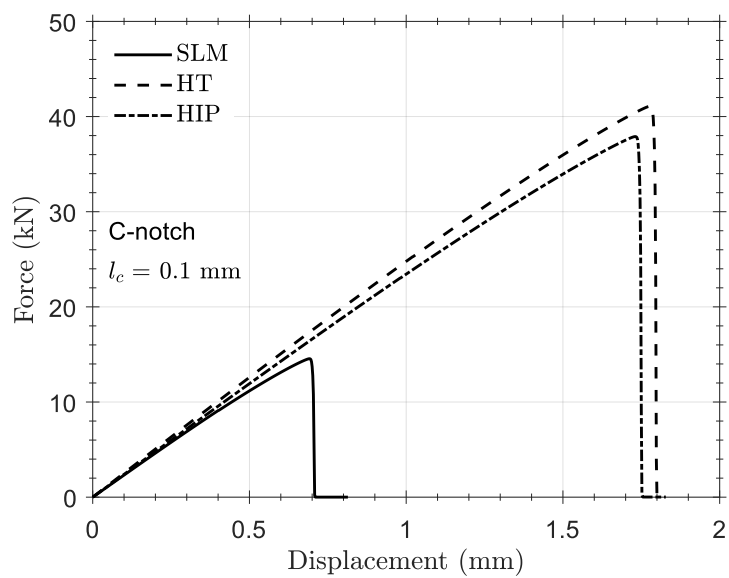

*Figure 11. Comparison of post-processing conditions for the C-notch and elastic behaviour.*

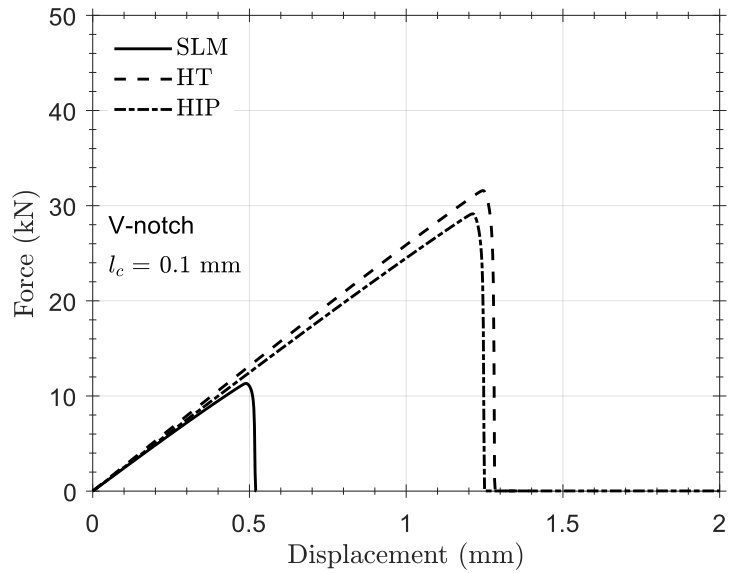

*Figure 12. Comparison of post-processing conditions for the V-notch and elastic behaviour.*

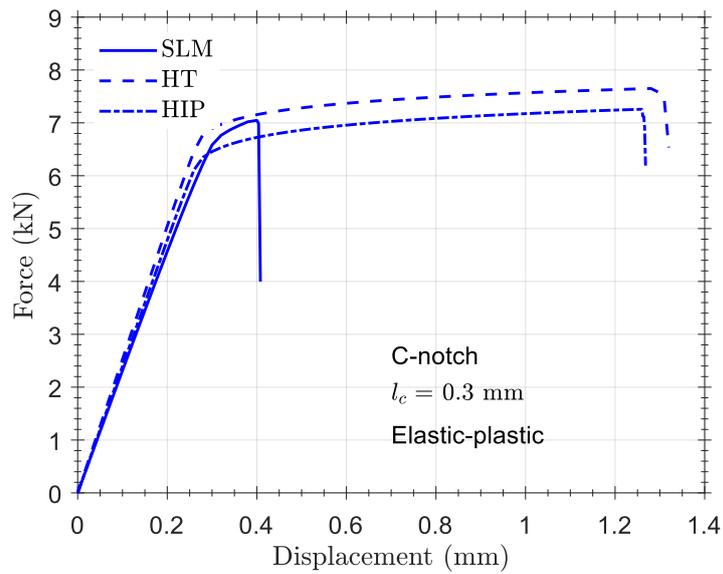

*Figure 13. Comparison of post-processing conditions for the C-notch and elastic-plastic behaviour.*

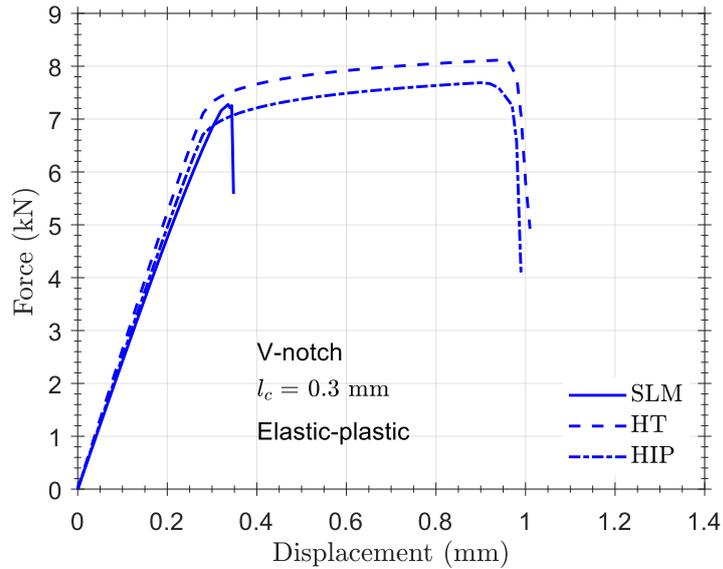

*Figure 14. Comparison of post-processing conditions for the V-notch and elastic-plastic behaviour.*

In contrast to the linear elastic assumption, plastic behaviour leads to convergence issues in the segregated step in which displacements are solved. For that reason, the failure of the whole ligament is not completely tracked in Figures 13 and 14, and only the first part of the load drop is obtained in numerical simulations. The minor difference between HT and HIP conditions in fracture behaviour cannot lead to any significant conclusion; actually, the slightly coarser microstructure for the HIPed samples in [16] should result in a higher toughness value. Nevertheless, the advantage of HIP over HT would be patent in a fatigue analysis, where the reduction in pore size is relevant; this is out of the scope of the present work.

### 4.3. Oxygen effects

Despite the great difference in the chosen oxygen fractions and partial pressures between the HT and HIP simulated non-inert environments, the boundary condition $C_s$ is similar in both conditions: 9.79 wt.% after HT and 8.18 wt. % after HIP. This similarity is caused by the much higher solubility at high-temperature HIP that counteracts the lower partial pressure. On the other hand, the oxygen-enriched layer is wider for the HIP post-process due to the much higher diffusivity, as shown in Figure 15. Nevertheless, it must be taken into account that two substantial simplifications have been assumed: (i) the oxide layer is not simulated and therefore the boundary condition for interstitial diffusion, i.e. $C_s$, is not reproducing the oxide/metal interface phenomena, and does not follow Sievert's law; (ii) solubility and diffusivity values have been fitted from a limited experimental data set from the literature [20] and thus the extrapolation to these temperature and pressure conditions might not be accurate enough.

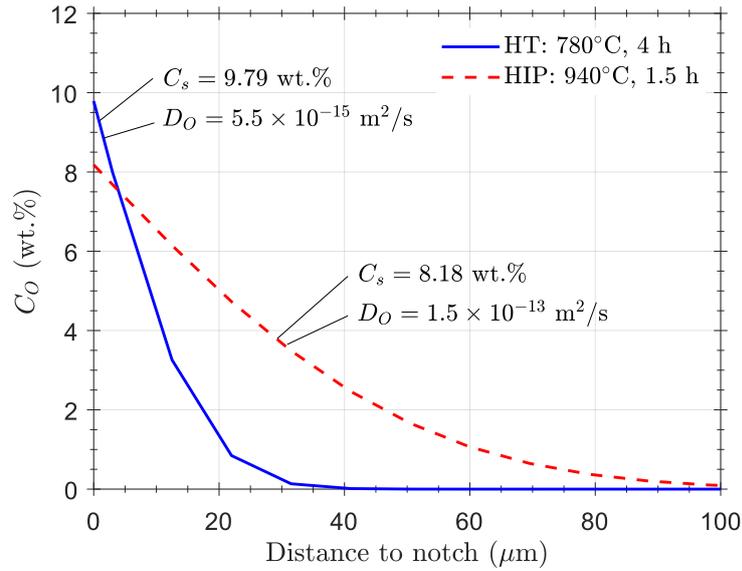

*Figure 15. Oxygen concentration in the notch plane for both HT and HIP post-processes.*

Figures 16 and 17 plot the effect of oxygen-modified mechanical properties, i.e. the concentration-dependent $K_c$ and $E$ values, on the fracture resistance of C-notch and V-notch geometries for the HIP conditions. For characteristic lengths of 0.1 and 0.3 mm and elastic behaviour, oxygen promotes a shift to lower forces and elongations at failures. However, this effect is more pronounced for the V-notch geometry due to the higher stress concentration that accentuates the oxygen-enhanced fracture energy reduction. It must be noted that an embrittlement coefficient $\chi = 0.01$ has been chosen in every oxygen-modified simulation and $\xi = 0.1$ is adopted for the calculation of the yield stress in plasticity. When the elastic-plastic material behaviour is implemented, the V-notch still shows oxygen-induced embrittlement, whereas the effect is negligible for the C-notch. This lack of oxygen influence on the C-notch fracture is expected when considering elastic-plastic behaviour since damage accumulation and final failure occur at the specimen centre, which is unaffected by oxygen. However, this finding contradicts the experimental fact that alpha case might promote failures at lower strains even in smooth specimens [49]. It must be noted that microcracking starting from the oxide layer has not been modelled here, so the stress concentration can be underestimated. In addition, as shown in oxygen distributions in Figure 15, the mesh density determined by Phase Field requirements, i.e. $l_e = l_c/7.5$, is coarse in comparison to the oxygen-enriched layer. For example, for $l_c = 0.1$ mm, the element size is $l_e = 13$ µm while the alpha case layer has a thickness of about 30 and 100 µm for the HT and HIP simulated conditions.

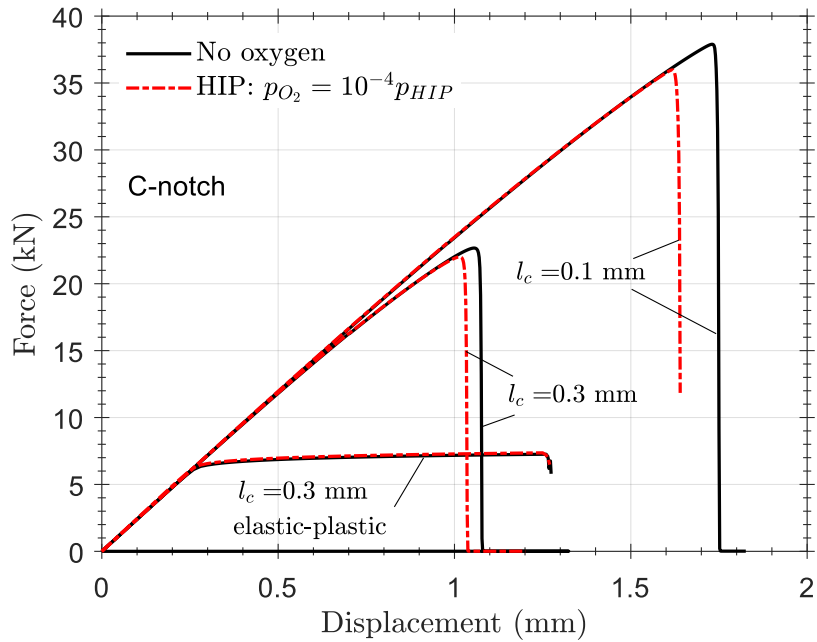

*Figure 16. Oxygen influence on C-notch fracture for a non-inert atmosphere during HIP.*

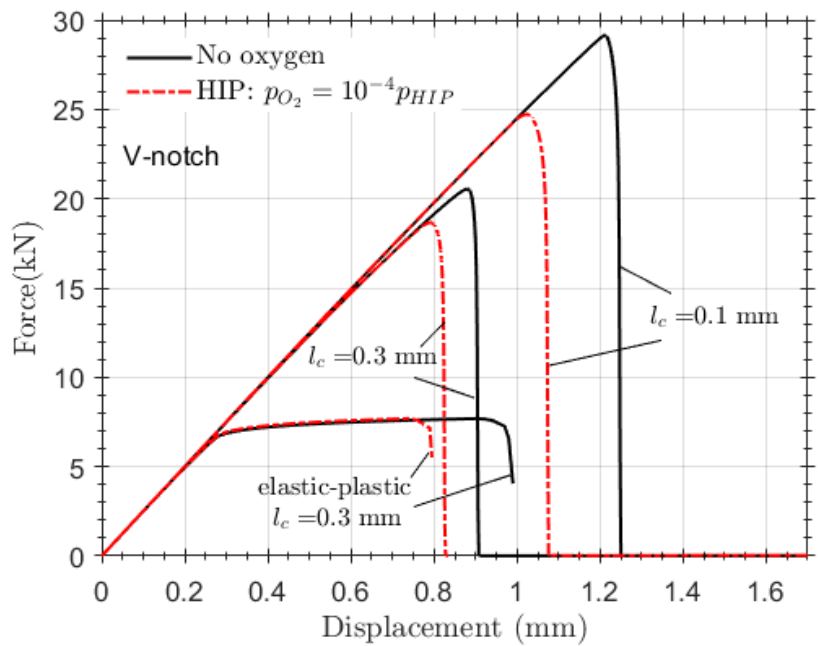

*Figure 17. Oxygen influence on V-notch fracture for a non-inert atmosphere during HIP.*

Comparing the effects of HIP and HT non-inert conditions, it is found that embrittlement is stronger for the HIP condition (Figures 18 and 19), as predicted by the thicker alpha case layer. Oxygen distribution after HIP at 940ºC for 1.5 hours, and the higher amount of oxygen in comparison to the 940ºC HT, suggests that alternative HIP cycles at lower temperatures and higher pressures should be beneficial for alpha case minimisation in the presence of oxygen impurities. As previously mentioned, this slight fracture difference could be well compensated by the HIP-enhanced improvement in fatigue life.

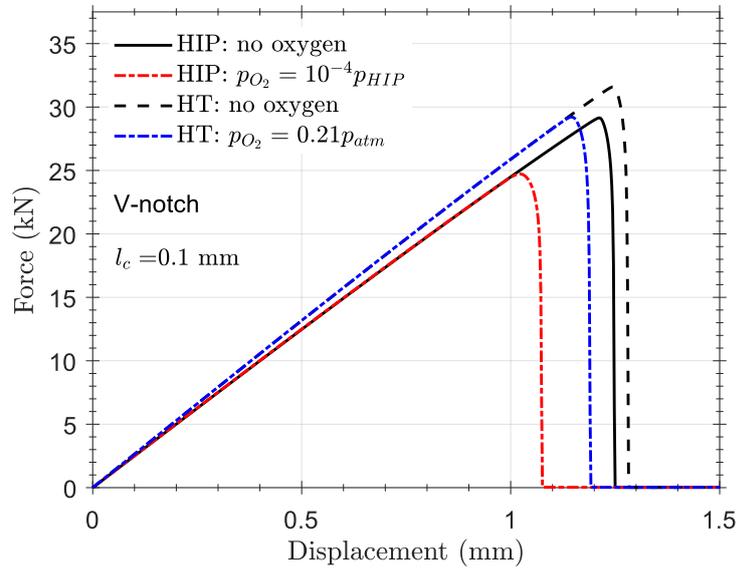

Figure 18. Oxygen influence on V-notch fracture for non-inert atmospheres during HIP and HT assuming elastic behaviour.

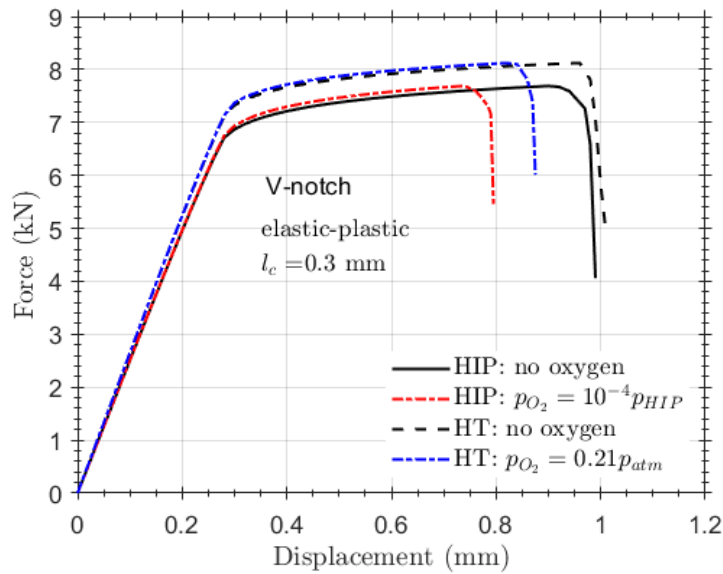

Figure 19. Oxygen influence on V-notch fracture for non-inert atmospheres during HIP and HT assuming elastic-plastic behaviour.

## 5. Conclusions

Phase Field modelling has been demonstrated to be a useful tool for predicting the notch fracture of additively manufactured Ti-6Al-4V specimens. Different post-processing conditions have been considered, with mechanical properties being extracted from literature data on samples produced by Selective Laser Melting [16]. The Phase Field fracture method is combined with elastic-plastic material behaviour and implemented into the commercial finite element software COMSOL. It is confirmed that a lower characteristic length results in failures at higher force and elongation levels; however, a reasonable choice of $l_c$ and $G_c$ for elastic-plastic material behaviour can produce unrealistic force values in a purely elastic simulation. Therefore, the present elastic-plastic modification of the Phase Field framework is essential to accurately model notch fracture after the toughness improvement by HT or HIP post-processes. Notch sensitivity has been assessed in two different geometries, C and V-shaped notches, finding that damage propagates from the notch surface in both geometries when the elastic behaviour is modelled, whereas the elastic-plastic response induces a plastic strain localisation at the specimen centre.

A two-step analysis has been considered to account for oxygen-informed mechanical behaviour. Oxygen diffusion during both heat treatment and hot isostatic pressing has been simulated and alpha case layers have been predicted for each condition. It has been found that the embrittlement effect of oxygen is more pronounced for the V-notch case, and it is negligible for the elastic-plastic behaviour in the C-notch since damage accumulates in the specimen centre. However, micro-cracking at the oxide layer has not been taken into account so the oxygen effect could be underestimated.

In future research, cyclic damage will be incorporated into Phase Field modelling to capture the HIP influence on fatigue behaviour. More efforts are also required to establish a consistent parameter choice strategy; the relationship between fracture energy, characteristic length values and the microstructure of Ti-6Al-4V after different thermo-mechanical processes will also be studied in more detail.

## Acknowledgments

The authors gratefully acknowledge financial support from the Junta of Castile and Leon through grant BU-002-P20, co-financed by FEDER funds. E. Martínez-Pañeda was supported by an UKRI Future Leaders Fellowship (grant MR/V024124/1). A. Díaz wishes to thank the Nanomechanical Lab of NTNU for providing hospitality during his research stay.